\documentclass[twocolumn,showpacs,showkeys,amsmath,amssymb,nofootinbib]{revtex4}

\usepackage{graphicx}% Include figure files
\usepackage{dcolumn}% Align table columns on decimal point
\usepackage{bm}% bold math

\begin{document}

\title{Spectroscopy of $^{18}$Na: Bridging the two-proton radioactivity of $^{19}$Mg}

\author{M. Assi\'e$^{1,2}$, F. de Oliveira Santos$^{1}$, T. Davinson$^{3}$, F. de Grancey$^{1}$, L. Achouri$^{4}$, J. Alc\'antara-N\'u$\tilde{\text{n}}$ez$^{5}$, T. Al Kalanee$^{4}$, J.-C. Ang\'elique$^{4}$, C. Borcea$^{6}$, R. Borcea$^{6}$, L. Caceres$^{1}$, I. Celikovic$^{7}$, V. Chudoba$^{8,9}$, D.Y. Pang$^{1}$, C. Ducoin$^{10}$, M. Fallot$^{11}$, O. Kamalou$^{1}$, J. Kiener$^{12}$, Y. Lam$^{13}$,  A. Lefebvre-Schuhl$^{12}$, G. Lotay$^{3}$, J. Mrazek$^{14}$, L. Perrot$^{2}$, A. M. S\'anchez-Ben\'itez$^{15}$, F. Rotaru$^{6}$, M-G. Saint-Laurent$^{1}$, Yu. Sobolev$^{14}$, N. Smirnova$^{13}$, M. Stanoiu$^{6}$, I. Stefan$^{2}$, K. Subotic$^{7}$, P. Ujic$^{1,7}$, R. Wolski$^{8,16}$, P.J. Woods$^{3}$}

\address{
$^1$ GANIL, CEA/DSM-CNRS/IN2P3, Caen France.\\
$^2$ Institut de Physique Nucl\'eaire Universit\'e Paris-Sud-11-CNRS/IN2P3 91406 Orsay France.\\
$^3$ School of Physics and Astronomy The University of Edinburgh Edinburgh EH9 3JZ UK.\\
$^4$ LPC/ENSICAEN, Blvd du Mar\'echal Juin 14050 Caen Cedex France.\\
$^5$ Departamento de Fisica de Part'iculas Universidade de Santiago de Compostela E15782 Spain.\\
$^6$ Horia Hulubei National Institute for Physics and Nuclear Engineering PO Box MG-6 76900 Bucharest Romania.\\
$^7$ VINCA Institute of Nuclear Sciences University of Belgrade P.0. Box 522 11001 Belgrade Serbia.\\
$^8$ Flerov Laboratory of Nuclear Reactions JINR RU-141980 Dubna Russia.\\
$^9$ Institute of Physics Silesian University in Opava Bezru\H{c}ovo n\H{a}m. 13 746 01 Opava Czech Republic.\\
$^{10}$ INFN - Sezione di Catania Via S. Sofia 64 Catania 95123 Italy.\\
$^{11}$ Subatech 4 rue Alfred Kastler BP 20722 F-44307 Nantes Cedex 3 France.\\
$^{12}$ CSNSM, Universit\'e Paris-Sud-11 CNRS/IN2P3 91405 Orsay-Campus France.\\
$^{13}$ CENBG Bordeaux France\\
$^{14}$ Nuclear Physics Institute AS CR 250 68 Rez Czech Republic.\\
$^{15}$ Departamento de F\'i'sica Aplicada Universidad de Huelva E-21071 Huelva Spain.\\
$^{16}$ The Henryk Niewodnicza\'nski Institute of Nuclear Physics PAS Cracow Poland.\\
}

%\date{}

%\date{\today}
\begin{abstract}
The unbound nucleus $^{18}$Na, the intermediate nucleus in the two-proton radioactivity of $^{19}$Mg, was studied by the measurement of the resonant elastic scattering reaction $^{17}$Ne(p,$^{17}$Ne)p performed at 4 A.MeV. Spectroscopic properties of the low-lying states were obtained in a R-matrix analysis of the excitation function. Using these new results, we show that the lifetime of the $^{19}$Mg radioactivity can be understood assuming a sequential emission of two protons via low energy tails of $^{18}$Na resonances.
\end{abstract}
\maketitle

%
% Introduction
%
From near to beyond the drip-lines, the nuclear force is no longer able to bind the interacting nucleons leading to instability of nuclei with respect to nucleon emission. On the proton-rich side of the chart of nuclides, the pairing force may lead to a situation where a drip-line nucleus is bound with respect to single proton emission but unbound to two-proton emission \cite{Gol60,2p}. Several types of two-proton emitters have been observed. On the one hand, there are the short-lived ($\tau_{1/2}\leq10^{-18}$ s) light nuclei such as $^{6}$Be, $^{12}$O or $^{16}$Ne  \cite{2p} where most probably the decaying ground state (g.s.) has large width so that it overlaps with 1p emitter states in the intermediate nucleus and the two protons are emitted mainly sequentially. On the other hand, there are the longer-living ($\tau_{1/2}\sim$ms) intermediate mass nuclei such as $^{45}$Fe, $^{54}$Zn or $^{48}$Ni \cite{tot} where the two protons could be emitted simultaneously. In all cases, it is essential to know the structure of the intermediate nucleus in order to understand the emission mechanism. But in many cases, it is very difficult to study experimentally the intermediate nucleus since it is located very far from the valley of stability. A new case of two-proton radioactivity was observed recently, this is $^{19}$Mg \cite{Muk07}. Its lifetime of 4.0(15) ps makes $^{19}$Mg an intermediate case between the short and long-lived nuclei. The measured lifetime and p-p angular correlations \cite{Muk08} are well described by the predictions of ref.\cite{Gri} when assuming only $d$-wave single-particle states in the low-lying structure of $^{18}$Na and thus a dominant $d^2$ single-particle configuration for $^{19}$Mg. Theoretical calculations of properties of $^{19}$Mg depend strongly on the assumption made about the structure of $^{18}$Na and its mirror nucleus. In this letter, we investigate both experimentally and theoretically the low lying spectrum of $^{18}$Na. We strongly refine the knowledge about the low-lying spectrum of $^{18}$Na and find that there are also low-lying $s$-wave states. These states should strongly boost the expected two-proton width.
\par
On the theoretical side, the structure of $^{18}$Na has been predicted assuming a core of $^{17}$Ne + proton structure \cite{Gri} or by coupling a neutron hole to the lowest states in $^{19}$Na \cite{For07}. In both cases, the low-lying structure of $^{18}$Na is found to have $d$-wave configuration. In the following, another theoretical approach is described and predicts also low lying s-wave states. The dimensionless reduced widths $\theta^2$ (sometimes called spectroscopic factors) where estimated with the shell model. The values shown in Tab. \ref{tab:spectro} were obtained with the OXBASH code \cite{oxbash} and the ZBM interaction \cite{zbm} in the 1p$_{1/2}$, 1d$_{5/2}$ and 2s$_{1/2}$ shells space. It predicts that the first six low lying states can be described mainly (with $\theta^2 >$ 0.5) by single particle configurations. The 1$^{-}_{1}$ and the 2$^{-}_{2}$  states are well described by the coupling $|^{17}\text{Ne}^{*}_{(3/2^{-})}\rangle\otimes |\pi \text{d}_{5/2}\rangle$. The 2$^{-}_{1}$ and the 3$^{-}_{1}$ states arise from the coupling $|^{17}\text{Ne}_{(1/2^{-})}\rangle\otimes |\pi \text{d}_{5/2}\rangle$. The two states 0$^{-}_{1}$ and 1$^{-}_{2}$ are described by the coupling $|^{17}\text{Ne}_{(1/2^{-})}\rangle\otimes |\pi  \text{s}_{1/2}\rangle$. The energies of the resonances can be predicted accurately from the known analog states of the mirror nucleus $^{18}$N \cite{nndc} according to the following prescription. We assumed pure core + neutron configurations to describe the states in $^{18}$N. A nuclear Woods-Saxon potential was fitted in order to reproduce the binding energy of the states in $^{18}$N. The obtained information was used to infer the position of the mirror core + proton states in $^{18}$Na taking into account the Coulomb interaction. The  states 0$^{-}_{1}$ and 1$^{-}_{2}$ are not known in $^{18}$N \cite{nndc}. In order to infer their position in $^{18}$N, we used the same method as in ref. \cite{For07}. We assumed the same mean value as for the $^{19}$O $\frac{1}{2}^{+}$ second excited state, i.e. $\sim$1.4 MeV \cite{nndc2} and an energy difference of 600 keV (that of the 2$^{-}$ and 3$^{-}$ couple in $^{18}$N). The partial proton widths were obtained using the relation $\Gamma_{i}=\theta^2\Gamma_{i}^{\text{W}}$ where $\Gamma^{\text{W}}$ is the Wigner limit. The calculated proton widths show very small contribution of the inelastic channel and some states are very narrow. A width of 22 keV is calculated for the ground state, an even smaller value of 1.3 keV is obtained if the spectroscopic factor given in the ref. \cite{Zer} is used. The spin-parity of the $^{18}$Na g.s. is predicted to be 1$^{-}$ and the separation energy S$_p$=1.3 MeV. Other mass extrapolation models \cite{audi} predicted separation energy for $^{18}$Na between S$_p$=1.5 MeV and S$_p$=1.9 MeV. These values mean that all states in $^{18}$Na are unbound to one-proton but also three-proton emission.
\par
 Experimentally, the $^{18}$Na nucleus was measured only once using a stripping reaction \cite{Zer,For05}, but the ground state was not clearly identified. Two peaks were observed, one with a proton separation energy of S$_p$=0.41(16) MeV and a width of $\Gamma$= 0.34(9)MeV, and the other with S$_p$=1.26(17) MeV and $\Gamma$= 0.54(13)MeV. If the first peak (S$_p$=0.41 MeV) corresponds to the ground state, then its position is in strong disagreement with the model predictions. Moreover, in this case, the $^{19}$Mg lifetime could not be understood since this nucleus would decay with an extremely fast ($\tau_{1/2}\leq$ $10^{-18}$ s) sequential emission of two protons through the intermediate $^{18}$Na ground state resonance. Even though the second peak seems more probable for the g.s.,  more experimental spectroscopic information is needed.
 \par
 Resonant elastic scattering is a powerful method to investigate the structure of unbound nuclei \cite{der} as it does not only provide the energies of the states but also widths and spins. However very few proton-rich unbound nuclei are accessible experimentally due to very low beam intensities as getting closer to the proton drip-line. The unbound nucleus $^{18}$Na, the intermediate nucleus in the two-proton radioactivity of $^{19}$Mg,  is one of the rare unbound nucleus to be accessible. In this letter we report on the measurement of the resonant elastic scattering reaction p($^{17}$Ne,p)$^{17}$Ne performed in inverse kinematics.
\par
\begin{figure}[h!]
\begin{center}
\includegraphics[height=5.5cm]{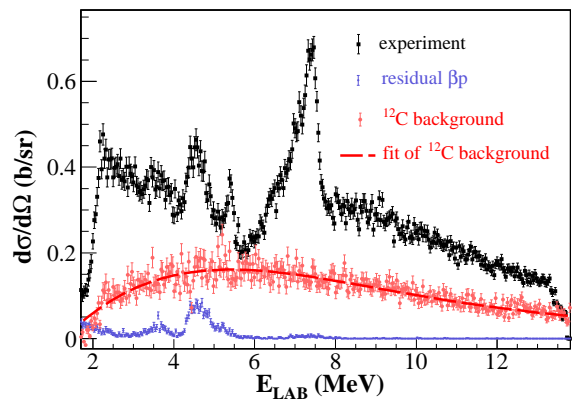}
\caption{(Color online) Proton spectrum measured between 5 and 20 degrees in the laboratory, reconstructed at 0 deg. The dashed red line represents a fit of the $^{12}$C background measured in this experiment. The dotted blue points represent the remaining contribution of $\beta$p decay of $^{17}$Ne after selection with ToF. }
\label{fig:Elab}
\end{center}
\end{figure}
\begin{figure}[h!]
\begin{center}
\includegraphics[height=6cm]{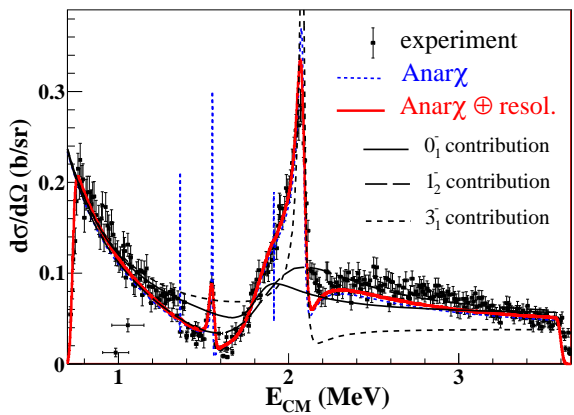}
\caption{(Color online) Measured excitation function of the $^{17}$Ne(p,$^{17}$Ne)p reaction presented as a function of CM energy. It is measured between 5 and 20 degrees (LAB) and reconstructed at 180 degrees in the CM.  Red and blue dotted lines show the R-matrix calculations based on the properties of the states given in Table \ref{tab:spectro}. The Red line is degraded by the energy resolution of the experiment, the blue dotted line is not. Three individual contributions are shown with black lines.}
\label{fig:results}
\end{center}
\end{figure}
\begin{figure}[h!]
\begin{center}
\includegraphics[height=4.2cm]{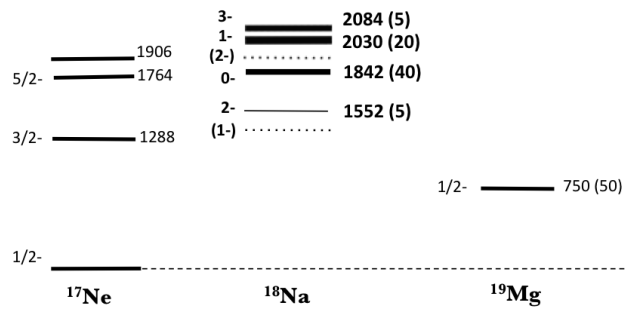}
\caption{States observed in $^{18}$Na and in the framework of $^{17}$Ne g.s. Decay energies are given in keV relative to the respective 1p and 2p thresholds. Dotted lines correspond to very narrow states that were not observed. The thickness of the lines is proportional to the width of the state. }
\label{fig:levels}
\end{center}
\end{figure}

%
% Experiment
%
The pure beam of radioactive $^{17}$Ne$^{3+}$ ions was produced by the Spiral facility at GANIL with a mean intensity of 10$^{4}$ pps and accelerated at 4 A.MeV. A beam of $^{17}$O$^{3+}$ ions was also produced in similar experimental conditions for calibration by comparison with the $^{17}$O(p,$^{17}$O)p reaction measured in direct kinematics \cite{o17}. The beam impinged on a fixed 50$\mu$m thick polypropylene C$_{3}$H$_{6}$ target coupled to a second rotating 50$\mu$m thick C$_{3}$H$_{6}$ target. The two targets together were thick enough to stop the $^{17}$Ne beam. This method, described for the first time in ref.\cite{der1}, has enabled us to measure the excitation function from 0.8 MeV to 3.8 MeV in the center-of-mass. Scattered protons were detected by a $\Delta$E-E annular telescope of silicon detectors placed at forward angles, called CD-PAD \cite{CDPAD}. The telescope was composed of a thin ($\approx$40$\mu$m) double-sided stripped silicon detector coupled to a 1.5mm thick silicon detector and was covering angles from 5 to 20 degrees in laboratory.
%\begin{figure}[h!]
%\begin{center}
%\includegraphics[height=4.8cm]{set-up2.eps}
%\caption{Sketch of the experimental set-up (see text for details).}
%\label{fig:set-up}
%\end{center}
%\end{figure}
%\begin{figure}[h!]
%\begin{center}
%\includegraphics[height=5.5cm]{ToF_prompt.eps}
%\caption{Time-of-flight versus energy deposited in the PAD detector spectra showing the kinematic line due to scattered protons and the vertical lines coming from $\beta$p decay.}
%\label{fig:ToF}
%\end{center}
%\end{figure}
The scattered proton spectrum had background from beta-delayed protons emitted in the beta decay of $^{17}$Ne. This nucleus decays with a lifetime of 0.109 s and a probability to emit protons of $\sim$90\%. More than 98\% of the $\beta$-delayed protons were rejected by using a 60 cm circular target (FULIS target \cite{fulis}) rotating at 1000 rpm. The ions were implanted in the target and moved away before their decay. A supplementary Microchannel Plate (MCP) was used for time of flight (ToF) and beam measurement with an efficiency close to 100\%. From ToF measurement and $\Delta$E-E selection, the scattered protons were identified and the proton spectrum was obtained (see Fig. \ref{fig:Elab}).
The residual $\beta$-delayed protons were subtracted following the same technique. The background produced by the presence of $^{12}$C in the target was measured using a pure carbon target with equivalent thickness and was also subtracted. Then, the proton spectrum was converted to the CM excitation function by using a Monte-Carlo algorithm taking into account the energy resolution of detectors and energy loss into the target. An overall energy resolution of 15 keV was obtained. In this analysis, it was assumed that protons were produced by elastic scattering reactions only.
%
%
% Results
%
\begin{table*}
\begin{ruledtabular}
\begin{tabular}{|c|c|cc|cc|c|c||c|cc|}
\textbf{J$^{\pi}$ }&\textbf{E$_{r}$}&\multicolumn{2}{c|}{ \textbf{$\theta^{2}\,^{17}$Ne$_{g.s.}$}}&  \multicolumn{2}{c|}{\textbf{$\theta^{2}$ $^{17}$Ne$^{*}_{3/2-}$}}&\textbf{  $\Gamma$ to $^{17}$Ne$_{g.s}$.} & \textbf{$\Gamma$ to $^{17}$Ne$^{*}_{3/2-}$}&\textbf{ E$_{r}$} &\textbf{$\Gamma$ to $^{17}$Ne$_{g.s}$}.&\textbf{$\Gamma$ to $^{17}$Ne$^{*}_{3/2-}$} \\
 &\textbf{(keV)} & \textbf{$1d_{5/2}$} & \textbf{$2s_{1/2}$}  & \textbf{$1d_{5/2}$} & \textbf{$2s_{1/2}$} & \textbf{(keV)}& \textbf{(keV)}& \textbf{(this work)}& \textbf{(this work)}&\textbf{(this work)}\\
\noalign{\smallskip}\hline
 \textbf{1$^{-}_{1}$}&1300& - & 0.086 &0.921& 0.183 &  22 (1.3)& 0& {-}& $<$ 1                &  $<$1 \\
\textbf{ 2$^{-}_{1}$}& 1500&0.644&  -& 0.311 & 0.042 &  8& 0 & 1552(5)   & 5(3)               & $<$1  \\
\textbf{ 0$^{-}_{1}$}& {\em1650}&- & 0.759  & - &  -&189 &  0&1842(40)   & 300(100)     & $<$10  \\
\textbf{ 2$^{-}_{2}$}&1950&0.004 & -   & 0.507 & 0.028 & 0.2& 0.45&  {-}     & $<$ 1         &  $<$ 1   \\
 \textbf{1$^{-}_{2}$}&  {\em 2450} &-  & 0.654  &0.031 & 0.027 & 1275 & 4.8& 2030(20)     &900(100)  & $<$100\\
\textbf{ 3$^{-}_{1}$}&2050&0.621&- &0.109 & - &  31& 0.04& 2084(5)    &42(10)            &    $<$ 1  \\
\end{tabular}
\caption{(Left side)Predicted resonances energies of the low lying $^{18}$Na states, deduced from analog states in the mirror nucleus $^{18}$N when they are known or deduced from the $^{19}$O 1/2$^{+}$ state (in italic), and dimensionless reduced widths $\theta^2$ calculated with the shell model and the corresponding proton widths to the g.s and the first excited state of $^{17}$Ne  (see text).(Right side) Measured spin, resonance energies and widths of the low lying states in $^{18}$Na.}
\label{tab:spectro}
\end{ruledtabular}
\end{table*}
Excitation function thus obtained (see  Fig. \ref{fig:results}) shows that the Rutherford scattering is dominant at low energy and it also shows at energies higher than 1.5 MeV several resonances reflecting the $^{18}$Na compound nucleus structure. Indeed, the position of these resonances is related to the energy of the excited states in the compound nucleus whereas their widths and shapes provide information on lifetimes (and spectroscopic factors) and spin-parity respectively. Spectroscopic properties of the low lying states in $^{18}$Na can be extracted using an R-matrix analysis of the measured excitation function. In order to perform this analysis, it is essential to find good initial conditions for the fit. For this, we used the properties predicted by the shell model (see Tab. \ref{tab:spectro}). A fit with the R-matrix code Anar$\chi$ \cite{anarki} was performed with energies and widths for the resonances as free parameters. The best fit obtained in this analysis is presented in Fig. \ref{fig:results}, with energies and widths of the resonances shown in Tab. \ref{tab:spectro}.
%
%
%
%Rutherford
The R-matrix fit agrees very well with the data in most part of the excitation function, including the low energy region where the cross section is mainly described by the Rutherford scattering. The first visible peak at resonance energy E$_{r}$=1.552(5) MeV is a very narrow with a measured width of 5(3) keV, and the shape of the peak agrees very well with a 2$^{-}$ spin assignment. These results are in excellent agreement with the predictions for the $2^-_1$ state (E$_{r}$=1.5 MeV and $\Gamma$ = 8 keV). The shape of the peak is not compatible with a 1$^{-}$ spin-parity. This first peak is followed by a broad peak which is well fitted when three resonances are taken into account. It corresponds to two broad states with spin-parity 0$^{-}$ and 1$^{-}$ and a narrower 3$^{-}$ state. This is in good agreement with the predictions for the 0$^{-}_1$, 1$^{-}_2$, 3$^{-}_1$ states. The 3$^{-}$ resonance is found very close to the predicted energy. The 1$^{-}_2$ resonance is down shifted by about 400 keV. We note the absence of a peak located at low energy and corresponding to the 1$^{-}$ ground state. As the energy resolution was of 12 keV in the CM, there are two possibilities: (i) either the 1$^{-}$ g.s. and the 2$^{-}$ first excited state of $^{18}$Na could not be resolved, this means that the energy difference between the two states is lower than 5 keV, (ii) or the 1$^{-}$ state is so narrow that it is not visible. The latter would be in agreement with shell-model calculations. Moreover, there may be a broad peak around 1.360 MeV which would be in agreement with the second peak of ref. \cite{Zer} but it could also correspond to some remnants of the $\beta$-p background. It could appear surprising to have such narrow states in an unbound nucleus located two steps beyond the proton drip-line. In fact, the escaping proton is kept longer inside the emitting unbound nucleus due to the Coulomb barrier, and also because of the structure of the state. Indeed the overlapping of $^{18}$Na g.s. and $^{17}$Ne g.s. is very small ($\theta^2$=0.086). Also, the 2$^{-}_{2}$ state is not visible in our spectrum, meaning that its width is very small, in agreement with the predictions. Zerguerras \emph{et al} \cite{Zer} observed a peak at $\Delta$M=24.19 MeV that was interpreted as an inelastic contribution from an excited state located at E$_{r}$=1.8 MeV. This interpretation is compatible with our results since the 0$^{-}$ state is located very close to this energy (E$_{r}$=1.82 MeV) and, within the uncertainties, the branching ratio for the inelastic channel could be as high as 15 \%. Few events of multiplicity 2 or more were observed. Our results are synthesized in the level scheme of Fig. \ref{fig:levels}.
\par
Lifetime of $^{19}$Mg for sequential two-proton decay through low energy tails of $^{18}$Na resonances can be estimated using the quasi-classical R-matrix type model from \cite{Gri07}. The main contributions arises through the broader resonances in $^{18}Na$, that are $\ell=0$ decays, so mainly from the 0$^{-}_1$ state ($\Gamma_{2}$=550 keV) and from the 1$^{-}_2$ state ($\Gamma_{2}$=500 keV). The calculated shell-model spectroscopic factors of $^{19}$Mg are: $\theta^2\left(|^{18}\text{Na}_{(0^-_1)}\rangle\otimes |\pi 2\text{s}_{1/2}\rangle \right)$= 0.0956 and $\theta^2\left(|^{18}\text{Na}_{(1^-_2)}\rangle\otimes |\pi 2\text{s}_{1/2}\rangle \right)$= 0.24. The estimated partial decay widths via separate isolated $s$-wave configurations are 2.4 10$^{-10}$MeV ($\tau$=1.9 ps) for the $0^{-}$ and 4.1 10$^{-10}$MeV ($\tau$=1.1 ps) for the 1$^{-}$. They exceed the total experimental width of $\Gamma_{exp}$=1.14 10$^{-10}$ MeV. The contribution of the 1$^{-}_1$ ground state resonance to the sequential decay is less than 1 \%.
It is clear that to explain the discrepancy the weights of the $s$-wave configurations in $^{19}$Mg should be suppressed compared to the weights obtained in the three-body calculations \cite{Gri} and in the shell-model calculations in this work (see $\theta_y=0.096$ and $\theta_y=0.24$). This implies strong domination of the $d$-wave configurations in the structure of $^{19}$Mg or/and configurations which can be interpreted as two protons plus excited states of $^{17}$Ne. Existence or nonexistence of the ground $1^{-}_1$ state not observed in this experiment does not influence this our conclusion.

%The sequential decay through tails of high-lying resonances 0$^{-}$ and 1$^{-}$ seems to represent a large contribution to the lifetime of $^{19}$Mg as suggested by \cite{For07}. However, the measured lifetime of $^{19}$Mg of 4.0(15) ps \cite{Muk07} is in good agreement with the calculation of  L.V. Grigorenko \cite{Gri} assuming a dominant $d^2$ configuration for $^{19}$Mg.  In this configuration, the predicted energies and widths of the $^{18}$Na states are in very good agreement with our results except for the 0$^{-}$ and 1$^{-}$ states. These states lie at much higher energies ($\sim$3.5 MeV) thus it reduces their contribution to sequential decay. In no way it is possible to fit the measured excitation function with 0$^{-}$ and 1$^{-}$ resonances located so high in energy.
%Another interesting comparison can be done with potential model results from ref. \cite{Gri}. With a potential based on the assumption that the d-wave is dominant, the energy and width of the states predicted are in very good agreement with our results except for the 0$^{-}$ and 1$^{-}_2$ states which lie at much higher energies ($\sim$3.5 MeV). However, the large Coulomb shift of the $\frac{1}{2}^{+}$ state in $^{19}$Na (located at 0.746(2) MeV) with respect to $^{19}$O $\frac{1}{2}^{+}$ state (located at 1.45MeV) indicates strong lowering of this doublet.

In conclusion, the measurement of the resonant elastic scattering reaction p($^{17}$Ne,p)$^{17}$Ne allowed us to obtain the properties of four low lying states in $^{18}$Na and suggests the presence of a very narrow ground state. The low-lying $s$-wave states observed imply important restrictions on the expected structure of $^{19}$Mg.  It favours the conclusion that the actual structure of $^{19}$Mg is strongly different from what is expected so far from different calculations. The sequential two-proton decay width of $^{19}$Mg nucleus via low energy tails of resonances corresponding to $s$-states in $^{18}$Na accounts fully for  the measured lifetime of $^{19}$Mg,  meaning that the decay of $^{19}$Mg is dominated by sequential emission.

We thank the GANIL staff, Dr. L. Grigorenko and the FULIS collaboration for their help. We acknowledge the support of the European Commission within the Sixth Framework Programme through I3-EURONS (contract no. RII3-CT-2004-506065), the support from the French-Romanian collaboration agreement IN2P3-IFIN-HH
Bucharest n$^{\circ}$03-33, the LEA COPIGAL, the R\'egion Normandie and the support from the UK STFC. We acknowledge the support of the French-Serbian CNRS/MSCI collaboration agreement (No 20505).
 

\begin{thebibliography} {99}
 \bibitem{Gol60} V.I. Goldansky, { Nucl. Phys.},  {\bf 19} (1960) 482.
 \bibitem{2p}B.Blank and M.P{\l}oszajczak, { Rep. Prog. Phys.},  {\bf 71} (2008) 046301 and references therein.
% \bibitem{6Be} D. F. Geesaman {\it et al}, { Phys. Rev. C},  {\bf 15} (1977) 1835.
% \bibitem{12O} R. A. Kryger {\it et al}, { Phys. Rev. Lett.},  {\bf 74} (1995) 860.
 \bibitem{tot} J. Giovinazzo {\it et al}, { Phys. Rev. Lett.} {\bf 89} (2002) 102501; M. Pf\"utzner {\it et al}, { Eur. Phys. J. A} {\bf 14} (2002) 279.; B. Blank {\it et al}, { Phys. Rev. Lett.} {\bf 94} (2005) 232501.; C. Dossat {\it et al}, { Phys. Rev. C} {\bf 72} (2005) 054315.
%
  \bibitem{Muk07} I.G. Mukha  {\it et al}, { Phys. Rev. Lett.},  {\bf 99} (2007) 182501.
   \bibitem{Muk08} I. Mukha  {\it et al}, { Phys. Rev. C},  {\bf 77} (2008) 061303(R).
  \bibitem{Gri} L.V. Grigorenko, I.G. Mukha and M.V. Zhukov, { Nucl. Phys. A}, {\bf 713} (2003) 372.
 \bibitem{For07}H.T. Fortune and R. Sherr,  { Phys. Rev. C}, {\bf 76} (2007) 014313.
  \bibitem{oxbash}B. A. Brown, A. Etchegoyen and W. D. M. Rae, { MSU-NSCL Report} No. {\bf 524} (1986).
 \bibitem{zbm} A.P. Zuker, { Phys. Rev. Lett.} {\bf 23} (1969) 983.
  \bibitem{nndc} D. R. Tilley, H. R. Weller, C. M. Cheves, and R. M. Chasteler, { Nucl. Phys. A} {\bf 595} (1995) 1.
 \bibitem{nndc2} ENSDF database, {\em http://www.nndc.bnl.gov/ensdf/}.
 \bibitem{Zer}T. Zerguerras {\it et al}, { Eur. Phys. J. A},  {\bf 20} (2004) 389.
  \bibitem{audi} G. Audi, A.H. Wapstra, { Nucl. Phys. A}, {\bf 595} (1995)  409; J. J\"anecke, P. Masson, { At. Data Nucl. Data Tables}, {\bf 39} (1988) 265; A. Pape and M. Antony, {\em At. Data. Nucl. Data Tables}, {\bf 39} (1988) 201.
 \bibitem{For05} H. T. Fortune  and R. Sherr, { Phys. Rev. C},  {\bf 72} (2005) 034304.
 \bibitem{der}K. Markenroth {\it et al}, { Phys. Rev. C},  {\bf 62} (2000) 034308.
 \bibitem{o17} J.C. Sens, S.M. Rafaei, A. BenMohamed and A. Pape, { Phys. Rev. C} {\bf 16} (1977) 2129.		
 \bibitem{der1} V.Z. Goldberg {\it et al}, { Phys. At. Nucl. }{\bf 60} (1997) 1186.
 \bibitem{CDPAD} A.N. Ostrowski {\it et al}, {\ Nucl. Instr. and Meth. A}, {\bf 480} (2002) 448.
 \bibitem{fulis} Ch. Stodel {\it et al}, Proceeding World Scientific EXON, Peterhof, Russia, July 2004, 180-187.
 \bibitem{anarki} E. Berthoumieux, B. Berthier, C. Moreau, J.P. Gallien and A.C. Raoux, { Nucl. Instr. and Meth. B}, {\bf 136-138} (1998) 55.
% \bibitem{ne17} M. Chromik {\it et al}, {\em Phys. Rev. C}, {\bf 66} (2002) 024313.
%\bibitem{lifetime} R.A. Kryger {\it et al}, { Phys. Rev. Lett.}, {\bf 74} (1995) 860.
\bibitem{Gri07}L.V. Grigorenko and M.V. Zhukov,{ Phys. Rev. C}, {\bf 76} (2007) 014009.
\end{thebibliography}
\end{document}